\documentclass[preprint]{ptephy_v1}%%%%%% to generate preprint number
\usepackage[bookmarks=false]{hyperref}% add hypertext capabilities
\usepackage{lineno}
\usepackage{amsmath} %for dealing with mathematics,
\usepackage{bm}

\usepackage{braket}
\usepackage[T1]{fontenc}

\begin{document}

\title{Reaction mechanism of quasi-free knockout processes in exotic RI beam era}

%%%% To generate auto affiliation numbers please use \author{}\affil{} command

\author[1,2]{Kazuki Yoshida}
\affil{Advanced Science Research Center, Japan Atomic Energy Agency, Tokai, Ibaraki 319-1195, Japan}
\author[2]{Junki Tanaka}
\affil{Research Center for Nuclear Physics, Ibaraki, Osaka 567-0047, Japan}

\begin{abstract}%
  The quasi-free nucleon knockout reaction has revealed the single-particle nature of nuclei.
  Thanks to the advances in experimental techniques and reaction theory, various new aspects of nuclei are being discovered using knockout reactions.
  In this article, we review the basic concept of the quasi-free knockout reaction, and recent achievements mainly at RIKEN using the liquid hydrogen target and detector system.
  We also present our new findings on the low-energy nucleon knockout reaction and the $\alpha$ knockout reaction. 
  The combination of the (microscopic) structure theory, reaction theory and experiments will be the key to a complete understanding of the $\alpha$ formation. 
  Although the $\alpha$ clustering phenomena have been studied intensively in the light mass region, its existence in the wide mass region of the nuclear chart is still unknown.  
  This is within the scope of the ONOKORO project, as well as noble clusters, e.g., $d$, $t$, and $^{3}$He. 
  The implementation of two-nucleon (or more) correlations in the reaction theory is essential to connect the properties of such clusters and the reaction observables. 
  A new framework is introduced for this purpose, which will also be applicable to the two-nucleon knockout reactions, e.g., $(p,3p)$, $(p,2pn)$, and $(p,p2n)$.
\end{abstract}

%\subjectindex{xxxx, xxx}

\maketitle

\section{Introduction}

An understanding of the quasi-free knockout reaction mechanism is essential to extract information about the internal structure of atomic nuclei from experimental observables. Such research has been conducted since the middle of the 20th century~\cite{Jacob66, Jacob73, Chant77, Chant83}, and progress has been made in understanding the structure of stable nuclei. Physical quantities observed in proton-induced knockout reactions targeting stable nuclei provided the excitation energy distribution of residual nuclei, the occupancy of nucleons in single particle orbits, and their momentum distributions. Recent research has indicated that the reaction cross sections of the $\alpha$ cluster knockout reaction are roughly proportional to the amount of $\alpha$ cluster formation on the nuclear surface, whose isotopic dependence of the reaction cross sections showed a good agreement with a theoretical prediction~\cite{Carey81, Carey84, Typel14, Yoshida16,Tanaka21, Yoshida22_Po}. Systematic cluster knockout reaction experiments are being performed to investigate the various clusters formed on the nuclear surface in ONOKORO project~\cite{Onokoro_en}. On the other hand, in research using RI beams, experimental techniques using inverse kinematics have improved, and many research results using quasi-free scattering of unstable nuclei have been reported~\cite{Santamaria15, Paul17, Flavigny17, Chen17, Lettmann17, Shand17, Olivier17, Cortes18, Liu18, Liu19, Paul19, Taniuchi19, Elekes19, Chen19, Cortes20, Cortes20_2, Sun20, Lokotko20, Lizarazo20, Frotscher20, Juhasz21, Juhasz21_V63, Browne21, Linh21, Koiwai22, Gerst22, Enciu22, Elekes22, Chen23, Linh24, Enciu24, Pengjie24} in the past decades from SEASTAR project~\cite{SEASTAR_webpage} at RIKEN RIBF, leading to progress in understanding unstable nuclear structures.

Inverse kinematics experiments using RI beams have several advantages: In proton-induced knockout reactions, a heavy ion beam is incident on a hydrogen target~\cite{Obertelli14} and RI nuclei can be studied. There are various types of knockout particles, including nucleons, such as protons and neutrons, and light ions, which are cluster particles. Since the residual nucleus travels straight with almost the same energy as the incoming beam in the laboratory frame, detailed analysis with spectrometers is possible. In addition, the emitted recoil protons have relatively high energy and can be measured. Measurement of these protons and knocked-out particles will complement the missing mass. 
Missing-mass spectroscopy is grounded in the conservation of four-momentum and provides a model-independent approach to reconstructing the mass of the unobserved residual system. In inverse kinematics proton-induced knockout reactions of the form $(p, pX)$, the missing mass $M_{\mathrm{miss}}$ is calculated as
\begin{align}
  M_{\mathrm{miss}} = \sqrt{ \left( E_b + m_p - E_p - E_X \right)^2 - \left| \vec{P}_b - \vec{P}_p - \vec{P}_X \right|^2 },
\end{align}
where $E_b$ and $\vec{P}_b$ denote the energy and momentum of the incident beam, and $E_p$, $\vec{P}_p$, $E_X$, $\vec{P}_X$ correspond to the energy and momenta of recoil proton and the knocked-out particle $X$, respectively. This formalism allows direct access to the excitation energy spectrum of the residual nucleus without requiring its detection, making it particularly suitable for theoretical comparisons and for isolating contributions from specific reaction mechanisms such as quasi-free scattering and clustering. It also facilitates the identification of discrete and continuum states, enabling a quantitative connection between reaction observables and nuclear structure models.
Reaction channels in which the final state decomposes into many particles can also be measured because the emitted particles are ejected forward with energy comparable to that of the incoming beams, and these measurements give an invariant mass. Due to these unique advantages of inverse kinematics, a wide variety of physical quantities can be measured, and the opportunities for more detailed reaction and structural research in RI beam research are increasing.

In theory, the knockout reaction is solved in the center-of-mass frame in practice. 
The laboratory frame, either the forward or the inverse kinematics, and the center-of-mass frame are connected by the Lorentz transformation in principle.
However, the observables depend on the statistics and hence on the kinematics. 
Kinematically more exclusive cross sections such as the triple-differential cross section can be obtained in the forward kinematics with stable nucleus target. 
On the other hand, it is usually difficult to have enough statistics in the inverse kinematics with RI beam, and the 1D longitudinal or transverse momentum distribution is often discussed in the inverse kinematics as in SEASTAR.
Since the knockout reaction is the three-body reaction, its kinematics and observables in the forward and the inverse kinematics, and their relation are rather complicated, as overviewed in a recent article~\cite{Uesaka24}.
It should be noted here that recently the triple-differential cross section of $(p,p\alpha)$ reaction has been measured and analyzed~\cite{Pengjie23}.
Also, the quadruple differential cross section has recently been proposed~\cite{Ogata23}, which brings us natural correspondence between the particle motion inside a nucleus and the shape of the cross section.

From the viewpoint of the reaction mechanism, from the 1970s through the 1980s, when the spallation reaction first produced RIs~\cite{Heckman72} and initiated the RI beam era~\cite{Tanihata85}, the reaction mechanism was simplified, and the momentum distributions are described by the Goldhaber model~\cite{Goldhaber74} to the zeroth approximation. This model assumes that $K$ nucleons out of $A$ are chosen at random to form a fragment having total fragment momentum $\bm{p}_{K}$. The relation between the mean square momentum of the fragment $\braket{\bm{p}_K^2}$ and that of the nucleus $\braket{\bm{p}_A^2}$ is derived and shown to agree well with a known phenomenological formula~\cite{Feshbach73,Bieser74}. Since then, the production cross section of incident nuclear spallation reactions has been systematically studied from a macroscopic perspective, and an empirical code for calculating the production cross section using the Glauber model, evaporation model, and intranuclear cascade model~\cite{Metropolis58, Metropolis58_2,Bertini63, Chen68,Yariv79, Kodama82,Sanchez17} has been developed. 
This method has been applied to systematic studies of reaction cross sections and transmutation of nuclear fuel waste. This framework treats knockout reactions as the classical multiple collisions of nucleons. On the other hand, in nuclear structural research, we aim to elucidate reaction mechanisms from a more microscopic perspective, focusing on reaction channels. Inclusive measurements have become popular with the rise of measurement instruments such as MINOS that support inverse kinematics using RI beams, and accurate theoretical descriptions of reaction mechanisms are essential to explain experimental data. 
It should be also noted here that only a nucleon and an $\alpha$ can be regarded as an inert particle in describing the knockout reactions.
As for $\alpha$ particle, there is no bound excited state below its proton threshold and the first excited state is at 20.2~MeV, slightly above the proton threshold and slightly below the neutron threshold.
Therefore, little contribution from the $\alpha$ excited states is expected through the channel couplings.
In other words, observed $(p,p\alpha)$ events can be safely regarded as quasi-free $\alpha$ knockout reaction.
Novel clusters such as deuteron, triton, and $^{3}$He, etc., have to be treated as fragile particles, and the coupling and transition between their ground state and excited and continuum states during the reaction process.

In this paper, we review the knockout reaction mechanism from a microscopic perspective. 
Section~\ref{sec:theory} explains the knockout reaction theory. 
Next, in Sec.~\ref{sec:nucleon_knockout} the nucleon knockout reactions are reviewed. 
We will explain the momentum distribution and its asymmetry obtained in experiments using the RI beam and the MINOS device, using the reaction mechanism introduced in Sec.~\ref{sec:theory}.  
In Sec.~\ref{sec:alphaKnockout}, recent researches for probing the $\alpha$ particle formation in nuclei ranging from light to heavy mass regions are introduced.
The reduced $\alpha$ decay width, which is essentially the $\alpha$ amplitude on the nuclear surface, is a key quantity to characterize the $\alpha$ decay.
If the $\alpha$ knockout reaction of heavy nuclei is only sensitive to the surface region by the absorption effect, the $\alpha$ knockout reaction allows us to deduce the reduced $\alpha$ decay width from the $\alpha$ knockout cross section, independently from the Coulomb barrier penetration process of the $\alpha$ decay.
Furthermore, in Sec.~\ref{sec:Beyond}, we will go beyond the standard nucleon and $\alpha$ knockout reactions and discuss two-proton knockout reactions and deuteron knockout reactions. 
To quantitatively connect the information of the deuteron-like $pn$ pair and correlated $pp$ pair formed in a nucleus and knockout cross section of them, it is essential to describe the complex reaction mechanism and theoretical progress has to be made.
This article is part of a series articles~\cite{Taniuchi_York, Yang_York, Chen_York, Kahlbow_York, Cortes_York, Liu_York, Tanaka_York, Bertulani_York} (expected to be) published in the MINOS Special Section.

\section{Quasi-free scattering inside a nucleus}
\label{sec:theory}
\begin{figure}[htbp]
  \centering
  \includegraphics[width=0.5\textwidth]{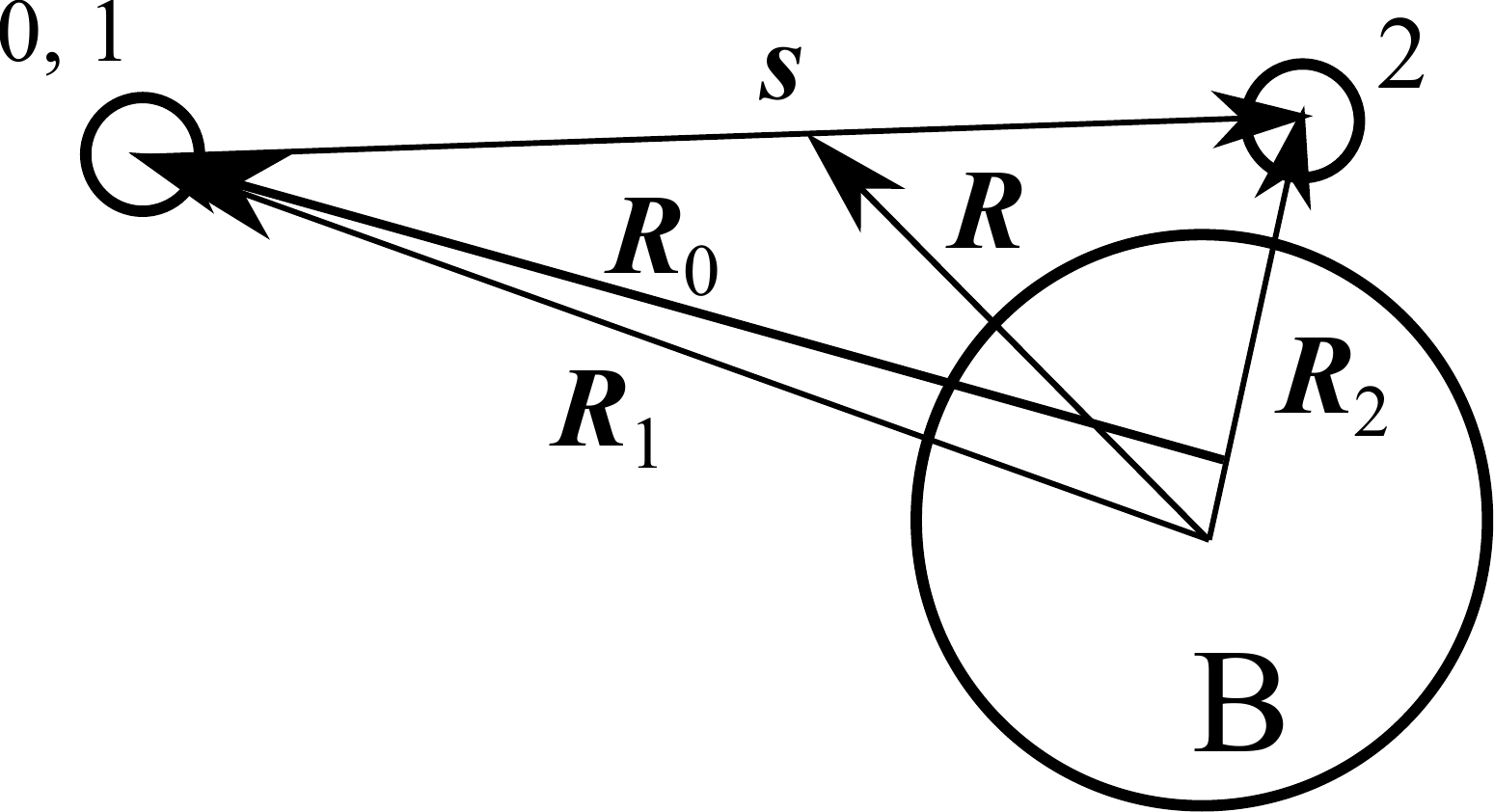}
  \caption{Coordinate and label for the particles.}
  \label{fig:mechanism-coordinate}
\end{figure}

The proton-induced nucleon knockout reaction, $(p,pN)$, is described as a three-body reaction of the incident proton, struck nucleon $N$, and the core (residual) nucleus B.
The target nucleus A is regarded as a bound state of $N$ and B.
The incident proton, outgoing proton, and struck $N$ are labeled as particles 0, 1, and 2, respectively.
The coordinates between these particles are defined as shown in Fig.~\ref{fig:mechanism-coordinate}.

\subsection{Impulse approximation}
In this section, the essence of the knockout reaction is introduced, along with the triple differential cross section, the quadruple differential cross section, and the momentum distributions.
Those quantities are essential ones in describing the knockout reaction cross sections, and in this review they are introduced using the distorted-wave impulse approximation (DWIA).
Based on the multiple scattering theory~\cite{Watson53,Francis53,Riesenfeld56}, the scattering of a nucleon from a nucleus can be described in terms of the multiple scattering between the incident nucleon and the nucleons in the target.
From this perspective, $(p,pN)$ reaction can be regarded as an extreme case where only a single nucleon inside a nucleus is involved in the reaction.
First of all, we recall the impulse approximation as a high energy approximation on the $NN$ collision inside a nucleus~\cite{Kerman59}, which supports our naive picture of the knockout reaction.  
In terms of the bare $NN$ interaction $v_{01}$ and the Green's operator $G$, the effective interaction $\tau_{01}$ of the $NN$ collision inside a nucleus A is given by 
\begin{align}
  \tau_{01} &= v_{01}\left(1+G\tau_{01} \right), \\
  G &= \left(E - H +i\varepsilon \right)^{-1}.
\end{align} 
Here, $H$ is not the free Hamiltonian but it contains the interaction between the bound nucleon and B, $V_{1\mathrm{B}}$.
In the nucleus A rest frame, $G$ can be written as 
\begin{align}
  G &= \left(E_\mathrm{A} - T_0 - T_1 - T_\mathrm{B} - V_{1\mathrm{B}} +i\varepsilon \right)^{-1},
\end{align} 
where $E_\mathrm{A}$ is the total energy in the A rest frame, $T_0$, $T_1$, and $T_\mathrm{B}$ are the kinetic energy operator of particle $0$, $1$, and B, respectively.
First, the impulse approximation assumes that $V_{1\mathrm{B}}$ is negligible compared to $E_\mathrm{A}$ when the incident proton has high kinetic energy.
In addition, since $T_\mathrm{B} \approx T_1 /\left( A-1 \right)$ in the A rest frame, $T_\mathrm{B}$ can be neglected when $A$ is large.
Thus, $G$ can be approximated as 
\begin{align} 
  G \approx G_0 = \left( E_\mathrm{A} -T_0 -T_1 +i\varepsilon\right)^{-1},
\end{align}
which is the free Green's operator of the $NN$ scattering in the free space.
This approximation results in the free $NN$ $t$-matrix as 
\begin{align} 
  \tau_{01} \approx t_{01} = v_{01}\left( 1 + G_0 t_{01} \right).
\end{align} 

\subsection{Distorted wave impulse approximation}
\label{sec:DWIA}
Based on the impulse approximation, the DWIA framework~\cite{Chant77,Chant83, Wakasa17} employs the $NN$ effective interaction as a transition interaction of the $(p,pN)$ reaction. 
Its transition matrix is given by
\begin{align} 
  T = \Braket{\chi_{1}\chi_{2}\Phi_{1}\Phi_{2}\Phi_{\mathrm{B}}|t_{01}|\chi_{0}\Phi_{0}\Phi_{\mathrm{A}}}.
  \label{eq:t-matrix}
\end{align}
Here, particle 0, 1, and 2, nucleus A and B refer to the initial proton, the emitted proton, the knocked-out nucleon, the target nucleus, and the residual nucleus, respectively.
$\chi_0$, $\chi_1$, and $\chi_2$ are the distorted waves between 0 and A, 1 and B, 2 and B, respectively.
The internal wave function of particles $(i = \mathrm{0, 1, 2, A, B})$ are denoted by $\Phi_{i}$.
Note that the indices for the spin, isospin, etc., are omitted from Eq.~(\ref{eq:t-matrix}) for simplicity.  
If the particle $i$ is a nucleon, $\Phi_i$ only represents its particle number, spin, and isospin.
To reduce the $1+2+\mathrm{B}$ three-body wave function in the final state into a product of the two-body scattering wave, $\chi_1\chi_2$, the semiclassical approximation discussed in Sec. 6 of Ref.~\cite{Koshel76} is adopted; 
The operators in the coupling term are replaced by the eigenvalue, i.e., 
\begin{align}
  -\hbar^2\frac{\bm{\nabla_1} \cdot \bm{\nabla_2}}{m_\mathrm{B}}
  \approx 
  \hbar^2\frac{\bm{K}_1 \cdot\bm{K}_2}{m_\mathrm{B}},
\end{align}
where $\bm{K}_1$ ($\bm{K}_2$) is the wave number of particle 1 (2) and $m_\mathrm{B}$ being mass of the residual nucleus.
One sees that if $m_{\mathrm{B}}$ is infinity the coupling term vanishes and the approximation is valid.
In the plane-wave limit (PW), Eq.~(\ref{eq:t-matrix}) is reduced to a product of the $NN$ t-matrix element and the overlap wave function as 
\begin{align}
  T 
  &\xrightarrow[\mathrm{PW}]{}
  \Braket{\bm{\kappa}'| t_{01} | \bm{\kappa}}
  \tilde{\varphi}_{2}(\bm{k}_2),
  \label{eq:PWfactorized}
\end{align}
where $\bm{\kappa}$ ($\bm{\kappa}'$) is the relative momentum between particle 0 and 1 in the initial (final) state and $\bm{k}_2$ is the momentum of particle 2 inside the target nucleus A, which can be approximately determined by the knockout reaction kinematics.
The overlap function $\varphi_{2}$ is defined as 
\begin{align} 
  \varphi_2 = \Braket{ 
                 \left[ \Phi_2 \otimes \Phi_{\mathrm{B}} \right]
                 |
                 \Phi_{\mathrm{A}}
               }
\end{align}
and $\tilde{\varphi}_{2}$ is that in the momentum space.
The triple differential cross section, which is the most kinematically exclusive cross section of the knockout reaction, is given by 
\begin{align}
  \frac{d^3 \sigma^{\mathrm{X}}}{dE_1^{\mathrm{X}} d\Omega_1^{\mathrm{X}} d\Omega_2^{\mathrm{X}}}
  &=
  \mathcal{J}_{\mathrm{XG}}F_{\mathrm{kin}}^{\mathrm{X}}
  \frac{(2\pi)^4}{\hbar v_\alpha}
  \frac{1}{(2s+1)(2j+1)}
  \sum_{\mathrm{spins}}
  \left| T \right|^2.
  \label{eq:tdx}
\end{align}
Here, $E_i^\mathrm{X}$ is the total energy of particle $i$ and $\Omega_i^\mathrm{X}$ is the solid angle of the emitted direction of particle $i$ defined in an arbitrary X-frame, and those without a superscript are defined in the center-of-mass frame.
$\mathcal{J}_{\mathrm{XG}}$ is the Jacobian from the center-of-mass frame to an arbitrary X-frame 
\begin{align}
  \mathcal{J}_{\mathrm{XG}}
  &=
  \frac{E_1 E_2 E_\mathrm{B}}{E_1^\mathrm{X} E_2^\mathrm{X} E_\mathrm{B}^\mathrm{X}},
  \label{eq:Jacobian}
\end{align}
and $F_{\mathrm{kin}}^\mathrm{X}$ is the phase volume.
$v_\alpha$ is the Lorentz invariant velocity of the incident particle.
The spin of the incident particle and the total angular momentum of the struck particle inside the target are denoted by $s$ and $j$, respectively.
Although indices are omitted, the $T$-matrix in Eq.~(\ref{eq:t-matrix}) includes spin degrees of freedom.
The summation over all the spin degrees of freedom is taken in the right-hand-side of Eq.~(\ref{eq:tdx}).

It is seen from Eqs.~(\ref{eq:PWfactorized}) and (\ref{eq:tdx}) that in the plane-wave limit, the knockout cross section is proportional to $\left| T \right | ^2$ and it shows the momentum distribution of the struck particle inside the nucleus A.
In reality, the scattering waves are not the plane waves but the distorted waves generated by the optical potentials.
Thus the momentum distribution of the knockout reaction is also distorted.

There are three types of input for the DWIA calculation.
Firstly, the optical potentials for the distorted waves of the incident and emitted nucleons.
Various parametrizations of the nucleon optical potential are available, for example, the global fit by Koning and Delaroche~\cite{Koning03} and the parametrization by the Dirac phenomenology~\cite{Hama90,Cooper93,Cooper09}.
See \cite{Chant77,Chant83,Wakasa17,Ogata23} for details of DWIA framework for knockout reactions.
Secondly, the $NN$ effective interaction for $t_{01}$.
One of the simple and powerful effective interactions is proposed by Love and Franey~\cite{Love81,Franey85}, and it is implemented in a recently published DWIA code {\sc pikoe}~\cite{Ogata23}.
Lastly, the overlap function or the single-particle wave function $\varphi_2$ is an input for DWIA calculation.
One phenomenological way is to obtain $\varphi_2$ as a nucleon bound state wave function in a Woods-Saxon shaped potential~\cite{Woods54}.
The optimal radius and the diffuseness parameters can be determined from the experimental momentum distribution of the $(p,pN)$ reaction~\cite{Enciu22,Enciu24}. 
The spectroscopic factor (SF) of the single-particle state can be discussed by comparing the theoretical and experimental knockout cross sections.
Also, theoretical predictions of the single-particle wave function or its SF are used in SEASTAR analysis, e.g., the large scale shell model~\cite{Shimizu_KSHELL}, valence-space in-medium similarity renormalization group (VS-IMSRG)~\cite{Stroberg19}, self-consistent Green's function (SCGF)~\cite{Soma20}, and Hartree-Fock-Bogolyubov (HFB) calculations~\cite{Bennaceur05}.

Various reaction theories are available other than DWIA.
One is the published computer code {\sc momdis}~\cite{Bertulani_momdis}, which employs the Glauber model with the scattering wavefunctions calculated in the eikonal approximation~\cite{Glauber_lecture,Hussein91}.
The Quantum-Transfer-to-the-Continuum (QTC) model~\cite{Moro15,Mario17,Mario18} describes the $(p,pN)$ reaction as a transfer reaction to the continuum state of the $NN$ pair, using the three-body wave function of the Continuum Discretized Coupled Channels method (CDCC)~\cite{Austern87}.
Also Faddeev-AGS method has been applied to $(p,pN)$ reactions~\cite{Crespo08,Crespo09,Crespo14,Cravo16}.
It has been confirmed by the benchmark comparisons~\cite{Mario20,Yoshidadwiatc} that these frameworks gives consistent results in many cases.

\subsection{Spectroscopic factor and its quenching}
As it is reviewed in Sec.~3.1. of Ref.~\cite{Aumann21}, many discussions have addressed the observability of the SFs and uncertainties on the extracted SFs using the nucleon knockout reaction.
Indeed, it has been shown that the occupation numbers (SFs) and momentum distributions of nucleons in nuclei are not served as observables from an effective filed theory perspective~\cite{Furnstahl2002}. 
However, debates on how unambiguously SFs can be extracted from reaction observables is still ongoing~\cite{Furnstahl10,Mukhamedzhanov10,Jennings11,Duguet15}. 
On top of the above discussion, it should be also noted here that compared to the $(e,e'p)$ reaction, more surface sensitivity is expected in $(p,pN)$ reaction due to stronger absorption effects by the nucleon-nucleus distorting potentials in both initial and final states. 
This effect hinders the contribution of the single-particle wave function in the internal region to the $(p,pN)$ cross section.
Comparison between the extracted SFs from $(e,e'p)$ and $(p,2p)$ are summarized in a review~\cite{Wakasa17}.

About $\sim 30$--$40\%$ reduction of SFs is expected to be originating from the missing correlations in the shell model~\cite{Lapikas93}, e.g., a spread of single-particle strength of deep orbitals, short-range and high-momentum correlations, etc.
A systematic trend of the quenching as a function of the proton-neutron binding-energy asymmetry has been reported by a systematic analysis of the nucleus induced one-nucleon removal reactions~\cite{Gade08,Tostevin21}.
Interestingly, different trends were found depending on the reaction types, $(e,e'p)$, nucleon transfer, $(p,pN)$ and nucleus induced nucleon removal reactions as summarized in Fig.~56. of Ref.~\cite{Aumann21}, suggesting that (some portion of) the trend originates from the reaction methods and adopted reaction theories, which is not yet clearly understood.

\subsection{Extension of DWIA to cluster knockout reactions}
The knockout reaction involving tightly bound clusters, e.g., the $\alpha$ cluster knockout reaction $(p,p\alpha)$, can be described in a similar manner as in $(p,pN)$ using Eqs.~\eqref{eq:t-matrix}--\eqref{eq:Jacobian}, but different inputs as follows.
Firstly, $\chi_2$ is replaced with the $\alpha$ distorted wave with respect to B. 
The global parametrization for the $\alpha$ optical potential~\cite{Nolte87,Avrigeanu94} or the double folding model~\cite{Egashira14,Furumoto23} can be used for this.
Secondly, instead of the $NN$ effective interaction, $p$-$\alpha$ effective interaction has to be used as $t_{01}$.
This can be prepared, for example, using the folding model~\cite{Toyokawa13} or $p$-$\alpha$ elastic scattering data, as is done in Ref.~\cite{Yoshida19}.
Lastly, $\varphi_2$ should be given as the overlap function between the $\alpha$ $+$ B system with the target nucleus A.
For this, various microscopic structure theories are available in light mass region, and also the mean-field type theory are recently extended to describe the $\alpha$ amplitude, as discussed in Sec.~\ref{sec:alphaKnockout}.

The knockout reactions of fragile clusters, e.g., the deuteron knockout reaction, is one of the possible probes for $p$-$n$ correlation in nuclei.  
The knockout reaction of such fragile nuclei cannot be described within the standard DWIA framework, and a new description combining the DWIA and the continuum-discretized coupled channels (CDCC) method is recently introduced~\cite{Chazono22} as discussed in detail in Sec.~\ref{sec:cdccia}.

\section{Nucleon knockout reaction}
\label{sec:nucleon_knockout}
\subsection{Momentum distribution and the single-particle orbit}
\label{sec:MomentumDistribution}
\begin{figure}[htbp]
  \centering
  \includegraphics[width=0.8\textwidth]{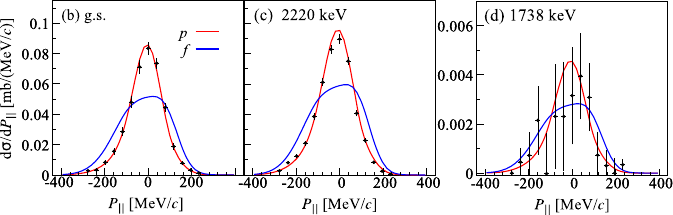}
  \caption{Longitudinal momentum distribution of $^{54}$Ca$(p,pn)^{53}$Ca reaction. (b) The $^{54}$Ca is in the ground state, (c) 2220 keV state (d) 1738 keV state.  Reprinted figure with permission from~\cite{Chen19} Copyright 2024 by the American Physical Society.}
  \label{fig:Chen54Ca}
\end{figure}

As discussed in Sec.~\ref{sec:DWIA}, the momentum distribution of the knockout cross section tells us the single-particle orbit of the struck particle, since the width of the momentum distribution becomes wider as the orbital angular momentum becomes larger.
A typical example is shown in Ref.~\cite{Chen19} and Fig.~\ref{fig:Chen54Ca}.
The longitudinal momentum distributions of $^{54}$Ca$(p,pn)^{53}$Ca reaction populating the $^{53}$Ca residue in the ground-state, 2220 and 1738 keV excited states are shown.
The ordering of the neutron orbitals in $^{54}$Ca is expected to be different from a traditional one due to the shell evolution driven by the lack of the tensor force between the neutron $1f_{5/2}$ and proton $1f_{7/2}$ orbitals in such neutron-rich nuclei~\cite{Otsuka01,Otsuka05}.
The comparison between the data and the DWIA curves strongly suggests that the valence neutron is in the $p$-wave, which supports the shell evolution in the $pf$ shell in the neutron-rich nuclei and the emergence of $N=34$ shell closure.

\subsection{Knockout reaction at low energy and the asymmetry of the momentum distribution}
\label{sec:Thomas_asymmetry}
It has been known that the longitudinal momentum distribution (LMD) of the nucleon removal reaction is asymmetric.
See Figs.~2 and 3 of Ref.~\cite{Gade05} for example.
In contrast, it has also been known that the LMD of the neutron halo removal reaction is almost symmetric (See Fig.~3 of Ref.~\cite{Nakamura14}).
The asymmetry of LMD has been theoretically discussed based on the DWIA framework~\cite{Ogata15}. 
The steep cut at the high momentum side is found to be due to the phase volume effect, which originates from the energy and momentum conservation law.
This effect becomes stronger as the removed (knocked-out) particle is tightly bound to the target.
The tail on the low momentum side of the LMD is enhanced by the momentum shift of the outgoing two nucleons in the interaction region by the attraction of the residue.  
Such final state interactions distort the scattering waves of the nucleons, thus local momenta of the scattering particles are different from their asymptotic ones. 
Therefore, the missing momentum of the knockout reaction in the reaction region differs from the asymptotic one and the momentum distribution of the knocked-out nucleon is also distorted.
From the above discussions, one may expect that the asymmetry is prominent when the separation energy is large and the beam energy is low, as theoretically predicted in Figs.~6 and 7 of Ref.~\cite{Ogata15} for $(p,pN)$ and experimentally seen in Figs.~2 and 3 of Ref.~\cite{Gade05} and Fig.~3 of Ref.~\cite{Nakamura14} for nucleon removal reaction with a nucleus target.

\begin{figure}[htbp]
  \centering
  \includegraphics[width=0.55\textwidth]{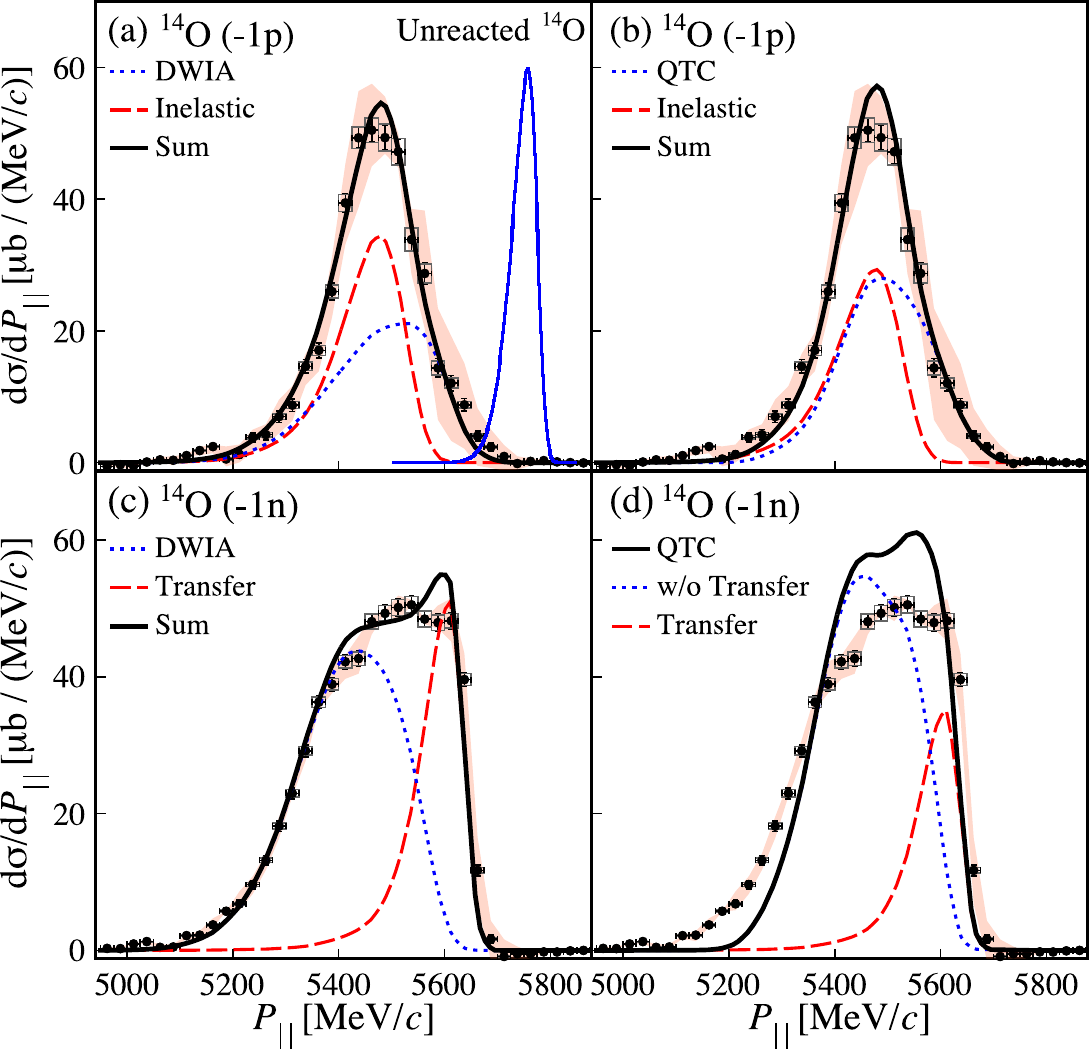}
  \caption{Longitudinal momentum distributions of $^{14}$O$(p,pn)^{13}$O and $^{14}$O$(p,2p)^{13}$N reactions at $\sim 100$ MeV/nucleon. Figure is taken from Ref.~\cite{Thomas23}.} 
  \label{fig:14O_ppn_PMD}
\end{figure}

Based on the above discussion, $^{14}$O$(p,2p)^{13}$N and $^{14}$O$(p,pn)^{13}$O were measured at $\sim 100$~MeV at RIBF, RIKEN~\cite{Thomas23}.
The proton separation energy ($S_p$) and the neutron separation energy ($S_n$) of $^{14}$O are $S_p = 4.2$~MeV and $S_n = 23.2$~MeV, respectively.
Due to the large gap in the separation energy, LMDs of the reactions show completely different shapes.
As seen in Fig.~\ref{fig:14O_ppn_PMD} (a) and (b), LMD of $^{14}$O$(p,2p)^{13}$N shows an almost symmetric shape, as expected from its small separation energy.
In contrast, $^{14}$O$(p,pn)^{13}$O shows a long tail in the low momentum side and a sharp cut in the high momentum side due to large $S_n$ as shown in Fig.~\ref{fig:14O_ppn_PMD} (c) and (d).
Note that a small tail in the low momentum side can be also seen in $^{14}$O$(p,2p)^{13}$N LMD due to the small beam energy.

The data were compared with calculations by DWIA and the Quantum Transfer-to-the-Continuum (QTC) calculations~\cite{Moro15,Mario17,Mario18,Mario20}.
In Fig.~\ref{fig:14O_ppn_PMD} (a) and (b), the DWIA and QTC results are shown in comparison with the $^{14}$O($p,2p)^{13}$N data, respectively. 
It was found that a contribution from the inelastic $^{14}$O($p,p')$ reaction followed by one-proton emission (dashed) is non-negligible and has to be added to the proton knockout cross section (dotted), as shown in Fig.~\ref{fig:14O_ppn_PMD} (a) and (b), to reproduce the experimental data.
It was also found in Fig.~\ref{fig:14O_ppn_PMD} (c) and (d) that the $^{14}$O$(p,d)^{13}$O transfer component is essential to reproduce the data.
It should be noted that the $^{14}$O$(p,d)^{13}$O cross section (dashed line in Fig.~\ref{fig:14O_ppn_PMD} (c)) has to be explicitly added to the DWIA result (dotted line) since the DWIA framework can only describe the knockout process.
On the other hand, both $^{14}$O$(p,pn)^{13}$O and $^{14}$O$(p,d)^{13}$O processes can be described in the QTC framework.
For comparison with the DWIA result, the QTC result is decomposed into $(p,d)$ and $(p,pn)$ cross sections in Fig.~\ref{fig:14O_ppn_PMD} (d).

In Ref.~\cite{Thomas23}, it has been clearly shown that asymmetry also exists in the LMD of the proton-induced knockout reaction, and it strongly depends on the nucleon separation energies.
It is also shown that the transfer and the proton decay contributions are negligible at around $\sim 100$~MeV incident energy. 
Since such contribution is negligible at higher energy around $250 A$~MeV as shown in Fig.~\ref{fig:Chen54Ca}, it will be interesting to investigate the incident energy dependence of such contributions in the nucleon removal cross sections.

\section{$\alpha$ knockout reaction}
\label{sec:alphaKnockout}
The proton induced $\alpha$ knockout reaction, $(p,p\alpha)$, has been utilized to probe the $\alpha$ clustering and its amplitude in a nucleus~\cite{Gottschalk70,Bachelier73,Bachelier76,Roos77,Landaud78,Nadasen80,Carey81,Carey84,Wang85,Nadasen89,Yoshimura98,Neveling08,Cowley09,Mabiala09}.
Recently, thanks to the experimental developments the triple differential cross section of the $(p,p\alpha)$ reaction can be obtained even in the inverse kinematics.

\begin{figure}[htbp]
  \centering
  \includegraphics[width=0.6\textwidth]{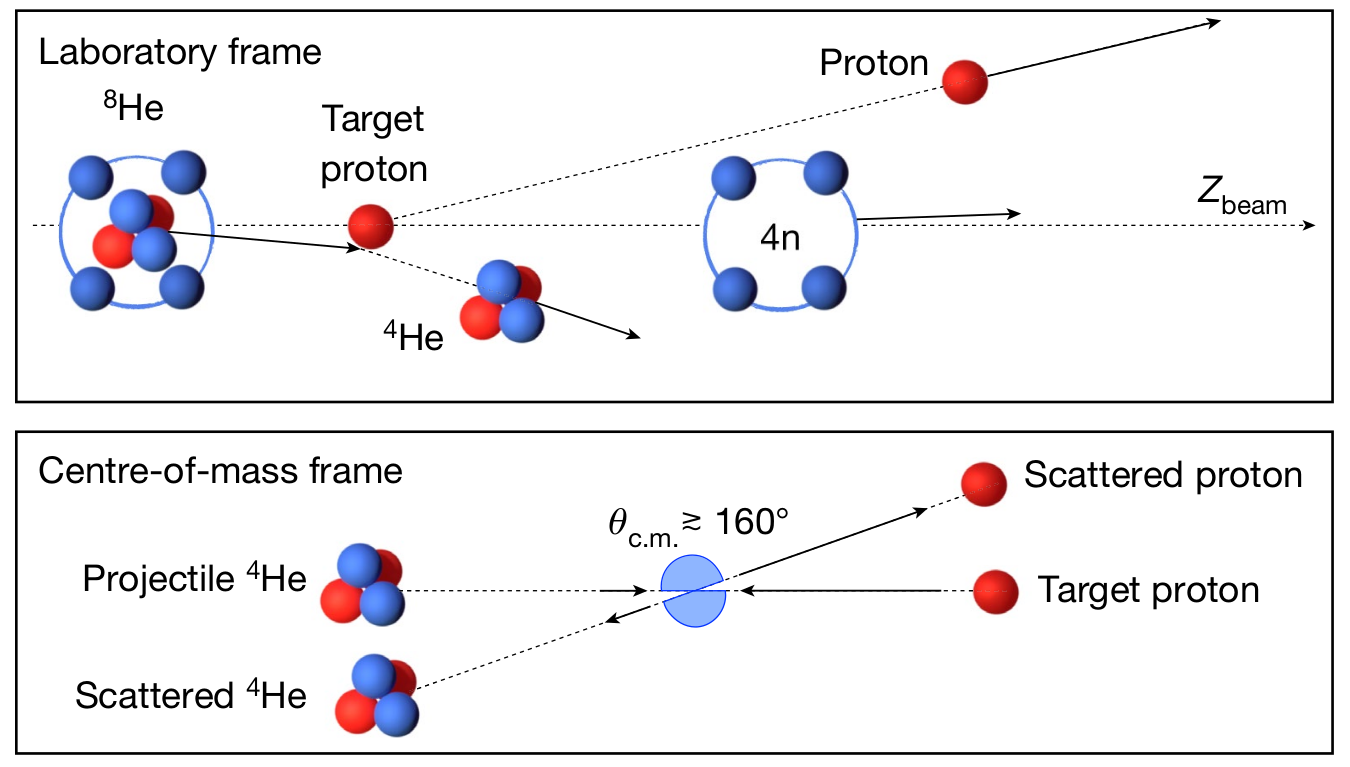}
  \caption{
    Schematic illustration of the $^{8}$He$(p,p\alpha)4n$ in the inverse kinematics.
    Top figure: Quasi-free $\alpha$ knockout reaction by the proton target in the laboratory frame.
    The $\alpha$ core of $^{8}$He is knocked out and the valence $4n$ is emitted to the forward direction.
    Bottom figure: The $p$-$^{4}$He elastic scattering in their center-of-mass frame.
    The backward angle scattering, $\theta_{c.m.} \gtrsim 160^\circ$, is considered in the experiment.
    Reprinted figure from~\cite{Duer22} Copyright CC BY 4.0.
    }
  \label{fig:Duer_kinematics}
\end{figure}

As the first $(p,p\alpha)$ experiment done with MINOS, the four-neutron resonance state was observed by the $^{8}$He$(p,p\alpha)4n$ reaction in the inverse kinematics~\cite{Duer22}.
The missing mass spectrum of the $4n$ system was reconstructed from the momenta of the charged particles and the energy and the momentum conservation law.
The quasi-free $\alpha$-$p$ scattering close to $180^\circ$ in the $\alpha$-$p$ center-of-mass frame was selected in the experiment (See Fig.~\ref{fig:Duer_kinematics}).
Choosing such backward angle $\alpha$-$p$ quasi-free scattering, the $\alpha$ core of $^{8}$He beam is slowed down by a large momentum transfer to the opposite direction of the beam.
Thus, the $\alpha$ core is knocked out from $^{8}$He nucleus without disturbing the valence $4n$ system, and they are emitted to forward angle with almost the same velocity as the beam, with almost no transferred momentum. 

\begin{figure}[htbp]
  \centering
  \includegraphics[width=0.6\textwidth]{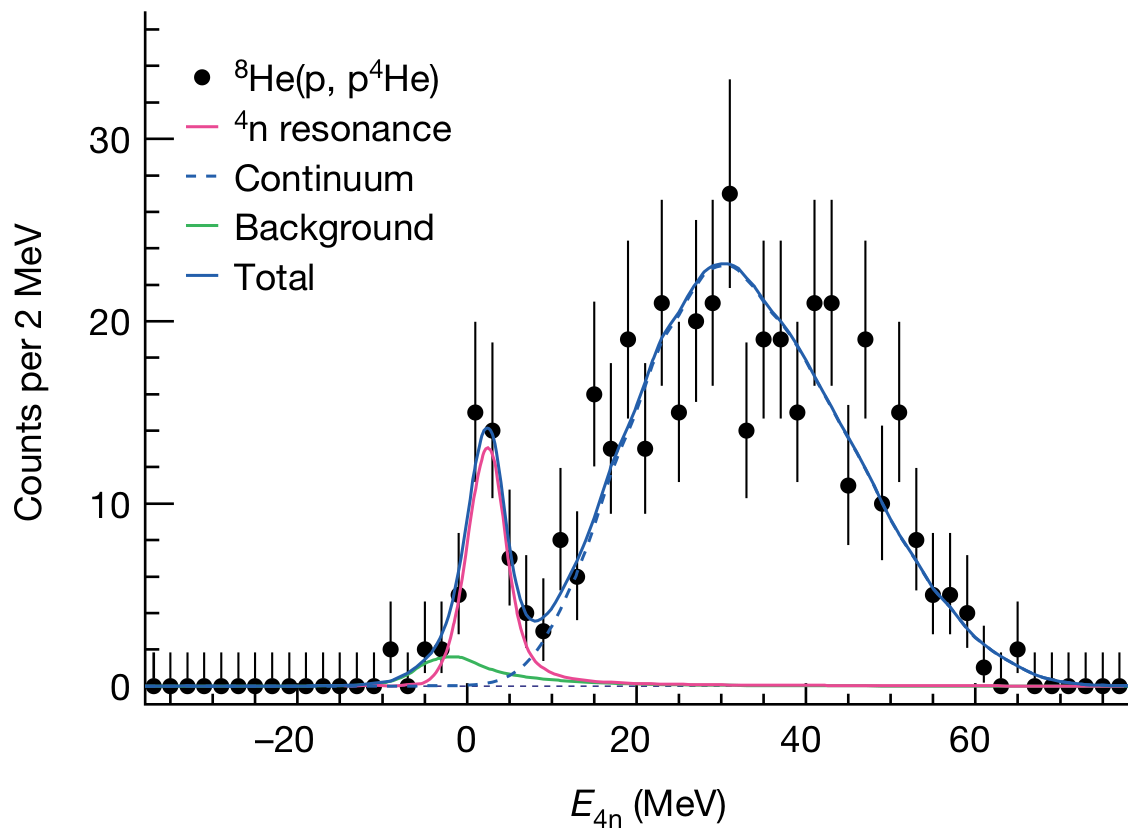}
  \caption{
    Observed missing mass spectrum of the $4n$ system in the final state of $^{8}$He$(p,p\alpha)4n$ reaction.
    Reprinted figure from~\cite{Duer22} Copyright CC BY 4.0.
    }
  \label{fig:Duer_spectrum}
\end{figure}

The missing mass spectrum of the $4n$ system obtained from the $^{8}$He$(p,p\alpha)4n$ reaction is shown in Fig.~\ref{fig:Duer_spectrum}.
A clear peak is observed above the threshold as shown in the pink line in Fig.~\ref{fig:Duer_spectrum}.
Assuming a Breit-Wigner shaped resonance state, the energy and the width are determined as $2.37\pm0.38$(stat.)$\pm0.44$(sys.)~MeV and $1.75\pm0.22$(stat.)$\pm0.30$(sys.)~MeV, respectively.
Detailed analysis and discussion of the $4n$ resonance state is discussed in other articles of this review.

\begin{figure}[htbp]
  \centering
  \includegraphics[width=0.50\textwidth]{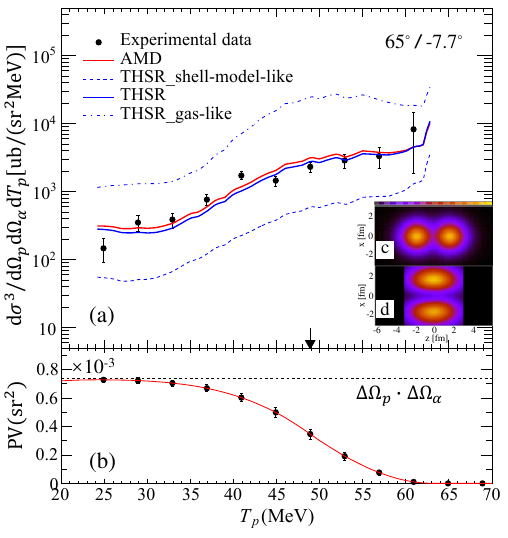}
  \caption{(a) The triple differential cross section of the $^{10}$Be$(p,p\alpha)^{6}$He reaction. 
    The experimental data are shown as a function of the proton emission energy $T_p$, in comparison with the DWIA calculations using the $\alpha$ amplitudes by THSR and AMD frameworks.
    (b) The corresponding phase volume of the reaction kinematics.
    (c) and (d) show the proton and the valence neutron density distributions predicted by THSR, respectively.
    Reprinted figure with permission from~\cite{Pengjie23} Copyright 2024 by the American Physical Society.
    }
  \label{fig:mechanism-Pengjie}
\end{figure}

The triple differential cross section of the $^{10}$Be$(p,p\alpha)^{6}$He has been reported in Ref.~\cite{Pengjie23}. 
The experimental data were compared with theoretical predictions of the DWIA calculations using the Tohsaki-Horiuchi-Schuck-R\"opke (THSR) wave function~\cite{Tohsaki01,Lyu16,Lyu18} and the antisymmetrized molecular dynamics (AMD) framework~\cite{Enyo15,Enyo16_1}.
Figure~\ref{fig:mechanism-Pengjie} shows the comparison between the experimental data and the theoretical results.
Both the THSR and AMD $\alpha$ amplitudes reproduce the data.
Also, as shown in the upper panel of Fig.~\ref{fig:mechanism-Pengjie}, artificial gas-like and shell-model-like compact $2\alpha$ structures of $^{10}$Be give larger and smaller cross sections, respectively, which indicate upper and lower extreme limits.
A similar analysis is ongoing for $^{12}$Be$(p,p\alpha)^{8}$He reaction and the difference of the $2\alpha$ structure in $^{10}$Be and $^{12}$Be will be studied by the $(p,p\alpha)$ reactions. 

\begin{figure}[htbp]
  \centering
  \includegraphics[width=0.5\textwidth]{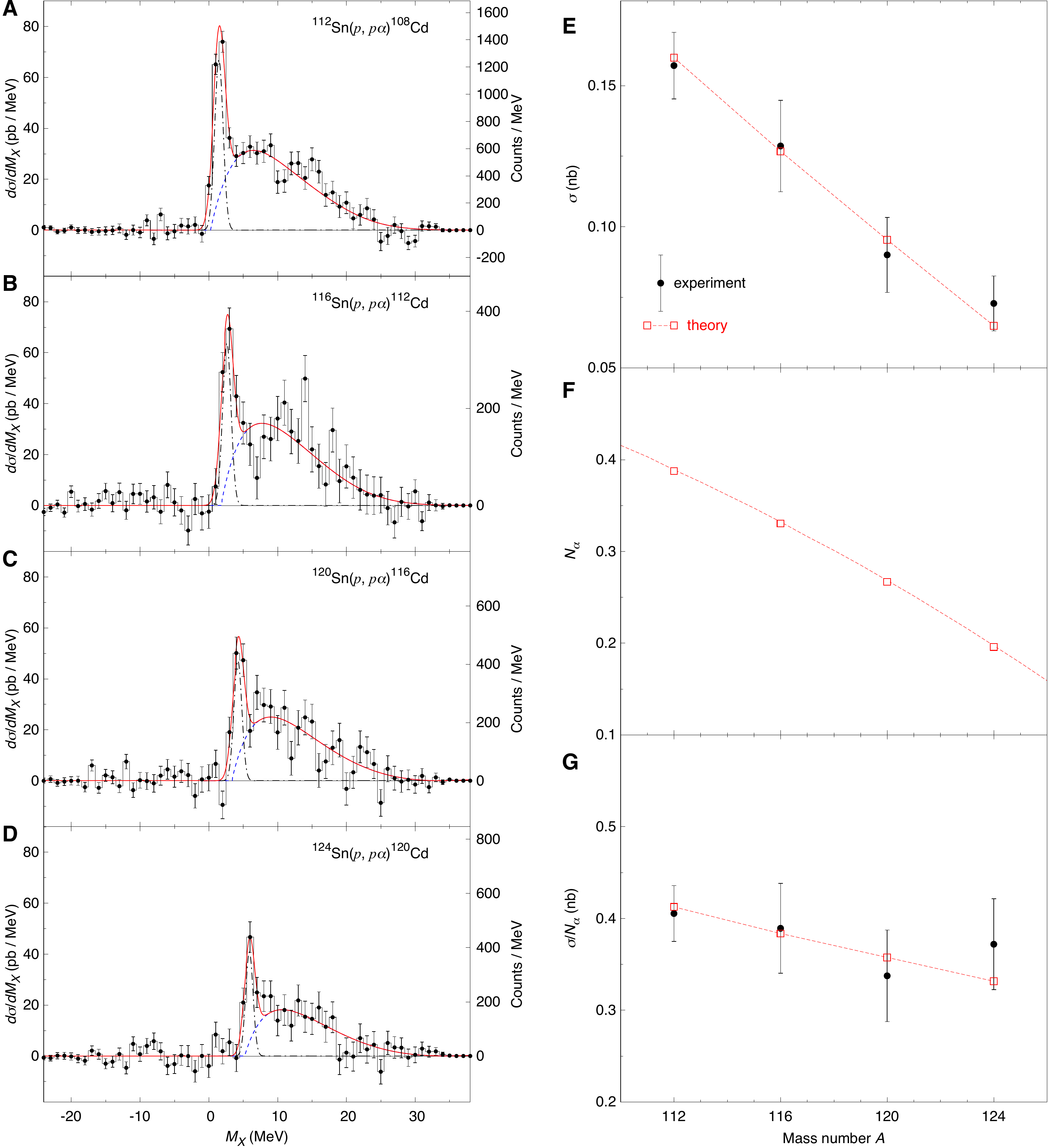}
  \caption{(A--D) Missing mass spectra of the $(p,p\alpha)$ reaction from $^{112}$Sn, $^{116}$Sn, $^{120}$Sn, $^{124}$Sn. (E) Experimental $\alpha$ knockout cross section (black circles with error bars) and theoretical cross sections (red square). (F) Theoretical prediction of the $\alpha$ particle numbers and its isotope dependence~\cite{Typel14}. (G) The ratio of the cross section to the $\alpha$ particle number. Reprinted figure with permission from~\cite{Tanaka21} Copyright 2024 by the The American Association for the Advancement of Science.
}
  \label{fig:Sn-ppa}
\end{figure}

 Our interest has been mainly focused on the $\alpha$ clustering in the light-mass region, but recently it has been extended across the nuclear chart.
Inspired by the theoretical prediction on the $\alpha$ formation on the surface of Sn isotopes~\cite{Typel14}, $\alpha$ knockout reaction from $^{112,116,120,124}$Sn were performed~\cite{Tanaka21}.
As shown in Fig.~\ref{fig:Sn-ppa} E, Sn$(p,p\alpha)$Cd cross sections and its isotope dependence were obtained, and a theoretical cross section based on the predicted $\alpha$ distribution~\cite{Typel14} reproduces the isotope dependence very well. 
The experiment confirmed the $\alpha$ particle formation on the surface of medium mass nuclei for the first time.
Also, the theoretical prediction of the suppression of the $\alpha$ particle formation due to the development of the neutron skin has been verified experimentally.

\begin{figure}[htbp]
  \centering
  \includegraphics[height=10\baselineskip]{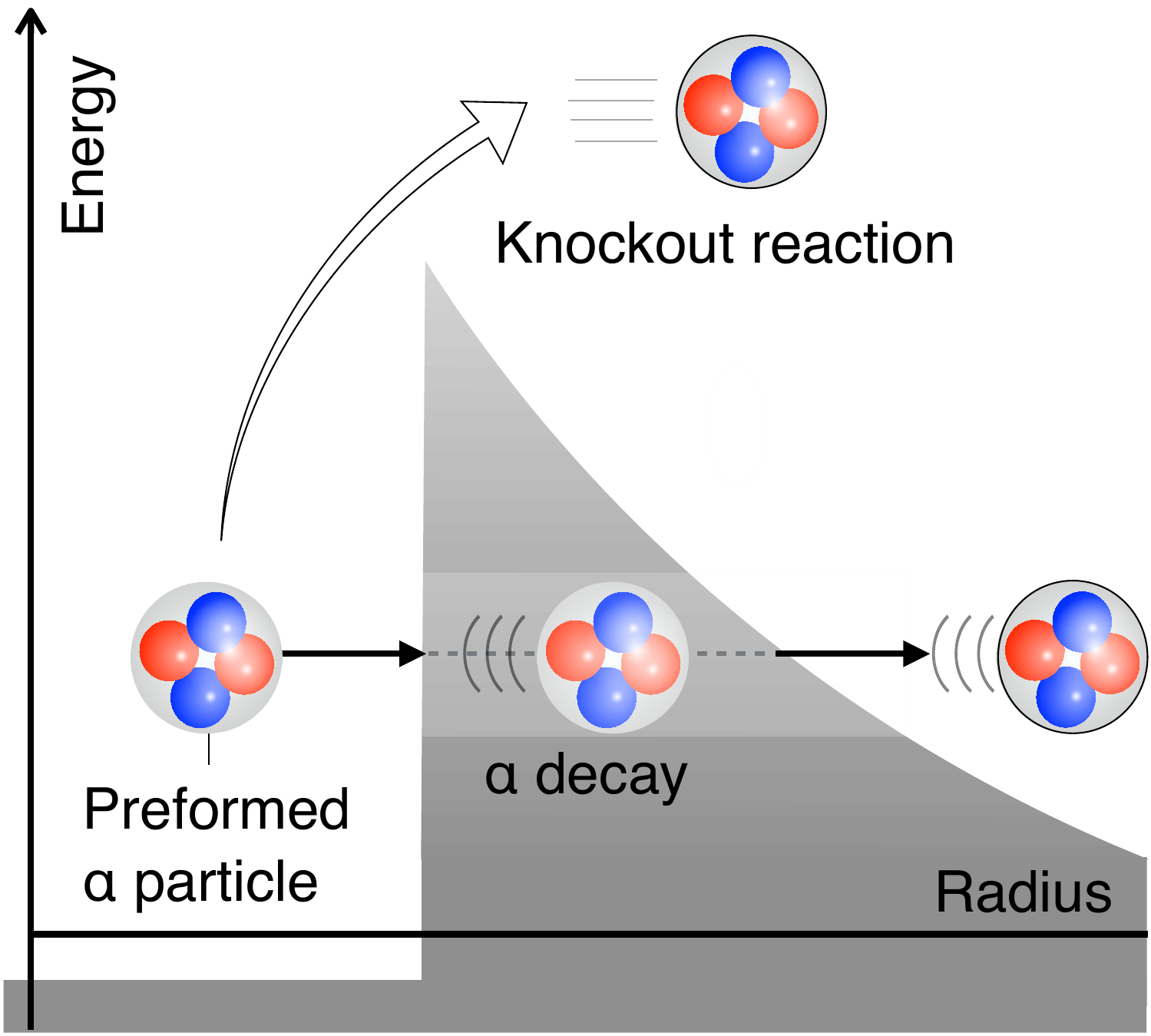}
  \includegraphics[height=10\baselineskip]{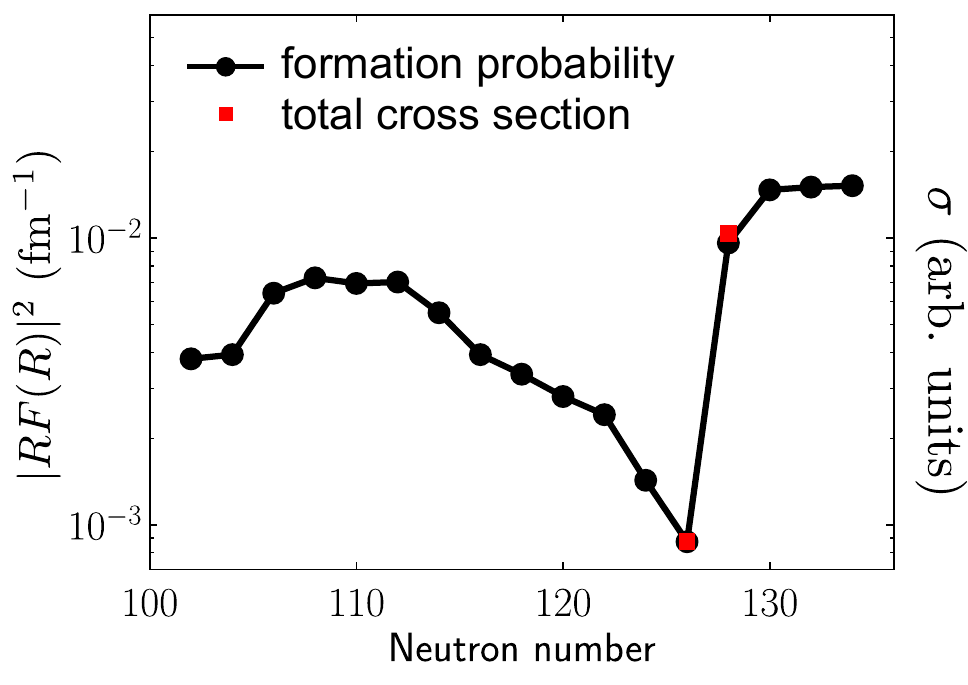}
  \caption{
    (Left) Illustrations of the $\alpha$ decay process and the $\alpha$ knockout reaction. The $\alpha$ decay is a process in which a preformed $\alpha$ particle penetrates the Coulomb barrier by quantum mechanical effect. In the $\alpha$ knockout process, a preformed $\alpha$ particle is hit by a high energy beam and knocked out with large energy and momentum transfer. The $\alpha$ particle has enough high energy to go above the Coulomb barrier and is less affected by the barrier.
    (Right) 
    A comparison between $\alpha$ formation probability of Po isotopes~\cite{Andreyev13} and the $\alpha$ knockout cross section of $^{210,212}$Po~\cite{Yoshida22_Po} using the $\alpha$ amplitude of Ref.~\cite{Qi2010}.
    Reprinted figure with permission from~\cite{Yoshida22_Po} Copyright 2024 by the American Physical Society.
  }
  \label{fig:Po_concept}
\end{figure}

The capability of the $\alpha$ knockout reaction for probing the $\alpha$ formation amplitude (reduced $\alpha$ width) on the surface of $\alpha$-decay nuclei has been investigated~\cite{Yoshida22_Po}.
The idea behind this work is that there is a large gap in the timescale of the $\alpha$ decay process and the $\alpha$ knockout reaction.
For example, a typical $\alpha$ decay nucleus, $^{212}$Po, in which we may expect an $\alpha$ + $^{208}$Pb component, has a lifetime of about $0.3~\mu s$.
On the other hand, the time scale of the knockout reaction is on the order of approximately $10^{-22}$ s ($\sim 30~\mathrm{fm}/c$), which is a typical timescale of direct reactions, which corresponds to the timescale of the incident particle traveling the diameter of a target nucleus.
This difference suggests that the $\alpha$ formation probability can be directly observed as $\alpha$ knockout cross section, independently of the $\alpha$ decay process, as shown in the left panel of Fig.~\ref{fig:Po_concept}.
In Ref.~\cite{Yoshida22_Po}, it has been also shown that $(p,p\alpha)$ reaction cross section is proportional to the reduced $\alpha$ width, which is also a key quantity of the $\alpha$ decay.
The left panel of Fig.~\ref{fig:Po_concept} shows a comparison between $\alpha$ formation probability of Po isotopes extracted from the decay half-life measurement~\cite{Andreyev13} and the $\alpha$ knockout cross section of $^{210,212}$Po using the theoretical $\alpha$ amplitude taken from Ref.~\cite{Qi2010}.
Although the cross section is given in the arbitrary unit, meaning that only $^{212}$Po/$^{210}$Po is reproduced, the work in Ref.~\cite{Yoshida22_Po} suggests that $(p,p\alpha)$ can be a good probe for the $\alpha$ formation in $\alpha$-decay nuclei.

The theoretical description of the $\alpha$ formation in heavy nuclei is challenging.
Recently, the $\alpha$-removal strength has been proposed from the mean-field approach ~\cite{Nakatsukasa23}.
At present, the theory assumes several approximations in order to make the numerical calculations feasible, e.g., the point $\alpha$ approximation as discussed in Sec. IIB1 of Ref.~\cite{Nakatsukasa23}.
However, as shown in Fig.~4. of Ref.~\cite{Nakatsukasa23}, the theory provides not only the $\alpha$-removal strength leaving the residue in the ground state, but also the strength going to the excited states of the residue.
Another new structure theory, a hybrid of the Hartree-Fock and AMD models, has recently been introduced~\cite{Kimura24}.
This model employs a set of localized Gaussians.
In contrast to AMD, the number of the localized Gaussians is not limited to the number of nucleons. 
This overcomes some difficulties in the AMD framework in the heavier mass region, and the binding energy, density distribution, etc. are improved as shown in Sec.3 of Ref.~\cite{Kimura24}.
These developments in the structure theories are essential for future studies of the $\alpha$ formation in the medium-mass nuclei and its systematics.

\section{Beyond the standard knockout reaction}
\label{sec:Beyond}
In this section, as a more advanced topic than the nucleon and $\alpha$ knockout reactions, we introduce two-nucleon knockout reactions, e.g., $(p,ppn)$, $(p,pd)$,  $(p,pnn)$, and $(p,3p)$. These reactions are expected to be a probe for $NN$ correlations in nuclei. To quantitatively discuss the correlated $NN$ pairs in nuclei, it is essential to understand the complex reaction mechanism and establish a theoretical framework that includes the fragile nature of the $NN$ pair and its dynamics in these reactions.

\subsection{nn correlation probed by knockout reaction}
\label{sec:nn_correlation}

In Ref.~\cite{Kikuchi16}, the dineutron correlation in Borromean nuclei $^{6}$He and the $^{6}$He$(p,pn)^{5}$He reaction were theoretically investigated, based on the $\alpha + n + n$ three-body model.
The theory predicted a large $nn$ correlation angle in the ground state of $^{6}$He in the momentum space; thus, it suggested a dineutron correlation in the coordinate space.
It was also found that the dineutron correlation is enhanced when focusing on the low $\alpha$-$n$ momentum region around $k_{\alpha\text{--}n} = 0.2$~fm$^{-1}$ of the $^{5}$He residue. 
In contrast, it was suggested that the cigar-like configuration could be seen in a high $\alpha$-$n$ momentum region around $k_{\alpha\text{--}n} = 1.0$~fm$^{-1}$. 
It was reported that the resonance state of the $^{5}$He$(3/2^{-})$ at $k_{\alpha\text{--}n} \sim 0.2$~fm$^{-1}$ drastically changes the relation between the dineutron correlation in $^{6}$He and the knockout reaction observables; it is important to choose a suitable knockout reaction kinematics to minimize the effect of the $^{5}$He resonance.
To avoid the effect from the resonance state, the authors of Ref.~\cite{Kikuchi16} suggested choosing off-resonant $\alpha$-$n$ relative momenta, $k_{\alpha\text{--}n} \sim 0.1$~fm$^{-1}$, to see the dineutron correlation by the knockout reaction.
The feasibility of the $^{6}$He$(p,pn)^{5}$He reaction using a high-intensity radioactive beam and the MINOS is also discussed in this paper.

\begin{figure}[htbp]
  \centering
  \includegraphics[width=0.5\textwidth]{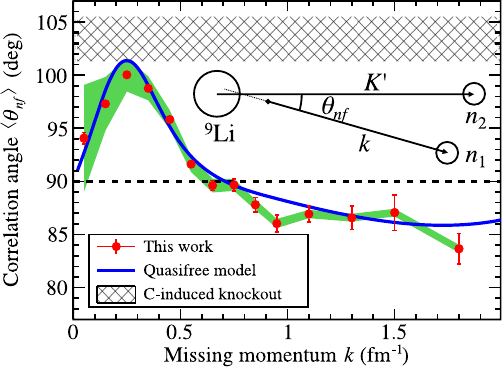}
  \caption{
    Mean value of the two-neutron correlation angle $\braket{\theta_{nf}}$ in the momentum space.
    The red points are the experimental data and the blue sold line is the quasi-free model result.
    Reprinted figure with permission from~\cite{Kubota20} Copyright 2024 by the American Physical Society.
  }
  \label{fig:11Li_dineutron}
\end{figure}

Building on the theoretical predictions from Ref.~\cite{Kikuchi16}, the localized dineutron formation in the $^{11}$Li nucleus was experimentally confirmed through the kinematically complete $^{11}$Li$(p,pn)^{10}$Li reaction~\cite{Kubota20}. 
This experiment, made possible by a high-intensity $^{11}$Li beam at RIBF and the MINOS, revealed a quasi-free knockout reaction with a momentum transfer larger than 1.5~fm$^{-1}$, significantly exceeding the two-neutron momentum in $^{11}$Li. 
The two-neutron $(1s)^2$, $(0p)^2$, and $(0d)^2$ configurations were effectively separated by selecting the missing momentum and the relative energy. 
In the missing momentum $k = 0.25$--$0.35$~fm$^{-1}$ region, the $s$- and the $p$-wave neutron components showed a similar magnitude, indicating an enhanced dineutron-like correlation in this region.
The mean two-neutron correlation angle $\braket{\theta_{nf}}(k) = \int \theta_{nf} P(\cos\theta_{nf},k) d(\cos\theta_{nf})$, with $P(\cos\theta_{nf},k)$ being a normalized $\cos\theta_{nf}$ distribution, is a key parameter in understanding dineutron correlation. 
In Fig.~\ref{fig:11Li_dineutron} (reprint from Ref.~\cite{Kubota20} Fig.~4), it is clearly shown that $\braket{\theta_{nf}}$ has a clear peak $\braket{\theta_{nf}} \sim 100^\circ$ at $k \sim 0.3$~fm$^{-1}$, indicating the presence of dineutron correlation in this region.
It is also found that $\braket{\theta_{nf}} \sim 90^\circ$ at $k \sim 0$, while it becomes less than $90^\circ$ in $k \gtrsim 0.6$~fm$^{-1}$ region.
A combination of the $^{9}$Li$ + n + n$ three-body model and the knockout reaction model of Ref.~\cite{Kikuchi16} reproduces the experimental data well as shown in the blue solid line in Fig.~\ref{fig:11Li_dineutron}.
This $k$ dependence of $\braket{\theta_{nf}}$ is consistent with the theoretical analysis of Ref.~\cite{Kikuchi16}.
The theoretical model suggests that the peak of $\braket{\theta_{nf}}$ at $k \sim 0.3$~fm$^{-1}$ can be interpreted as the dineutron correlation having a peak at $r \sim 3.6$~fm from the center of $^{9}$Li.

\subsection{Pairing effect on $(p,pN)$ cross sections}
\label{sec:Paul}
\begin{figure}[htbp]
  \begin{minipage}{0.47\textwidth}
    \centering
    \includegraphics[width=1.0\textwidth]{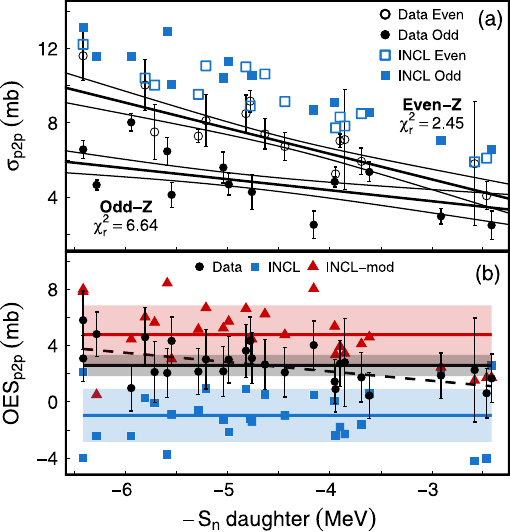}
    \caption{
      (a) Comparison between the inclusive $(p,2p)$ cross section (black circles) and the intranuclear cascade (INCL) calculations (squares) and
      (b) odd-even splitting of the cross sections,
      as the neutron separation energy $-S_n$ dependence.
      See Ref.~\cite{Paul19} for details.
      Reprinted figure with permission from~\cite{Paul19} Copyright 2024 by the American Physical Society.
    }
    \label{fig:paul_p2p}
  \end{minipage}
  \hspace{0.02\textwidth} 
  \begin{minipage}{0.47\textwidth}
    \centering
    \includegraphics[width=1.0\textwidth]{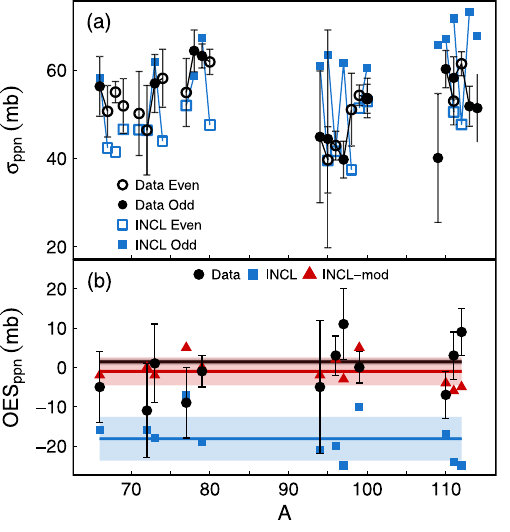}
    \caption{
      (a) Comparison between the inclusive $(p,pn)$ cross section (black circles) and the intranuclear cascade (INCL) calculations (squares) and
      (b) odd-even splitting of the cross sections,
      as the mass number dependence.
      See Ref.~\cite{Paul19} for details.
      Reprinted figure with permission from~\cite{Paul19} Copyright 2024 by the American Physical Society.
    }
    \label{fig:paul_ppn}
  \end{minipage}
\end{figure}

A systematic survey of the inclusive $(p,2p)$ and $(p,pn)$ cross sections from 55 neutron-rich nuclei was conducted in Ref.~\cite{Paul19}.
The measured inclusive $(p,2p$) cross sections are shown in Fig.~\ref{fig:paul_p2p} (a).
The even- (odd-) $Z$ projectiles are shown as open (filled) circles in Fig.~\ref{fig:paul_p2p} (a).
Both the $(p,2p)$ cross section from the even and odd $Z$ projectiles show a decreasing trend as $S_n$ decreases, but an odd-even effect can be seen;
the even-$Z$ projectiles give a larger cross section than odd-$Z$ projectiles for the same $S_n$ of the residue.
The Akaike information criterion (AIC)~\cite{Akaike74} analysis supports two separate linear trends with different slopes for even and odd $Z$ projectiles.
Figure~\ref{fig:paul_p2p} (b) shows the odd-even splitting (OES) of the cross section, $\mathrm{OES}_{p2p}$.
See Ref.~\cite{Paul19} for the definition of OES.
A constant regression (black solid line) and a linear regression with weak $S_n$ dependence (black dashed line) are suggested by a $\chi^2$ analysis.
The linear decreasing trend of the $(p,2p)$ cross sections can be explained as follows.
As nuclei become neutron rich, fewer bound states are below $S_n$ of the $A-1$ residual nuclei.
Therefore, the $(p,2p)$ cross section decreases as $S_n$ decreases.
The OES is related to the pairing interaction.
In $(p,2p)$ reactions on odd-$Z$ nuclei that leave even-$Z$ residues, the pairing interaction reduces the level density.
If the weak linear decreasing trend is true, such dependence can be understood by reducing the pairing as the neutron number increases, as suggested by the mass measurement~\cite{Tu90} and the predictions~\cite{Madland88,Pastore13}.

In contrast to the $(p,2p)$ case, the $(p,pn)$ cross sections do not show $S_n$ dependence as shown in Fig.~\ref{fig:paul_ppn} (a).
This is understood as follows.
As the neutron number increase along the isotopic chain, the neutron removal cross section is expected to increase.
However, as discussed in the $(p,2p)$ case, the $S_n$ of the residual nuclei decreases with the mass number, and hence the number of the bound states of the residue decreases as the neutron number increases.
No odd-even splitting was observed in the $(p,pn)$ case as shown in Fig.~\ref{fig:paul_ppn} (b).
This is because the reduced level density in the even-$N$ residues compensates for larger $S_n$ of them. 
Thus the total $(p,pn)$ cross sections leaving bound residues are similar in even-$N$ and odd-$N$ cases.
See Ref.~\cite{Paul19} for details and discussions on the comparison with the INCL calculations.

\subsection{Knockout reaction of fragile clusters and NN correlation}
\label{sec:cdccia}
\begin{figure}[htbp]
  \centering
  \includegraphics[width=0.8\textwidth]{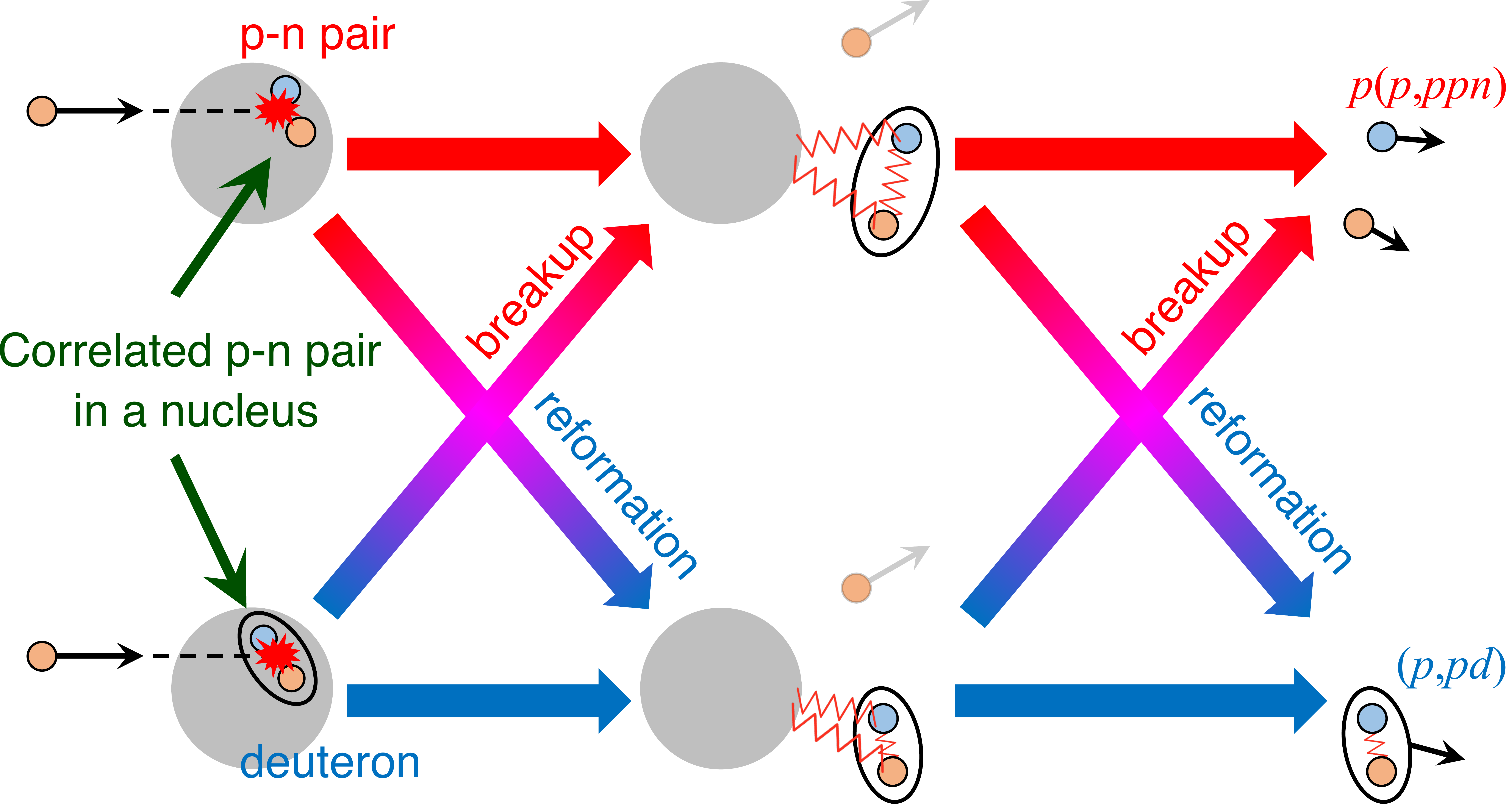}
  \caption{Possible reaction path of the correlated $p$-$n$ pair in a nucleus to the $p$-$n$ pair knockout reaction observables.}
  \label{fig:mechanism-CDCCIA}
\end{figure}

In contrast to the $(p,pN)$ and $(p,p\alpha)$ reactions, knocking out a fragile particle, such as deuteron, is not as straightforward.
The description of such reactions is theoretically challenging.
The deuteron knockout reaction, $(p,pd)$, can be a probe for the $p$-$n$ correlation in a nucleus.
Since the deuteron is fragile and its breakup and reformation processes play important roles in deuteron-involved reactions, the $(p,pd)$ reaction has to be described with a theory beyond the standard DWIA.
Recently, a combination of the DWIA and the continuum-discretized coupled channels (CDCC) method~\cite{Kamimura86,Austern87,Yahiro12} was recently developed~\cite{Chazono22}.
The framework, CDCCIA, contains two important features.
One is the description of the elementary process.
CDCCIA employs the $NN$ effective interaction as the transition interaction and the transition between the $p$-$n$ ground state (deuteron) and the continuum states in the elementary process is taken into account.
The other is the CDCC description for the $p$+$n$+$\mathrm{B}$ three-body wave function in the final state of the $(p,pd)$ reaction.
The transition between the $p$-$n$ ground and continuum states by the final state interaction is also taken into account in a CDCC manner.
These treatments allow us to connect the correlated $p$-$n$ pair to the $(p,pd)$ and $(p,ppn)$ reaction observables, as illustrated in Fig.~\ref{fig:mechanism-CDCCIA}.

Before the bombarded proton hits the $p$-$n$ pair, the $p$-$n$ pair is not necessarily the deuteron ground state.
Information about the $p$-$n$ pair in the target nucleus should be given from structure theory predictions, e.g., the two-nucleon amplitude, as an input to the CDCCIA calculation.
Once such input is given, the CDCCIA framework can describe the breakup and the reformation processes of the $p$-$n$ pair, by the elementary process and the final state interaction, as shown in Fig.~\ref{fig:mechanism-CDCCIA}.
The CDCCIA framework is not only applicable to $(p,pd)$ and $(p,ppn)$ but is also extensible to $(p,3p)$ and $(p,pnn)$ in a similar manner.
The only the difference would be whether the two-nucleon pair has the bound state or not.

\subsection{$(p,p2N)$ reaction and multiple scattering}
\begin{figure}[htbp]
  \centering
  \includegraphics[width=0.5\textwidth]{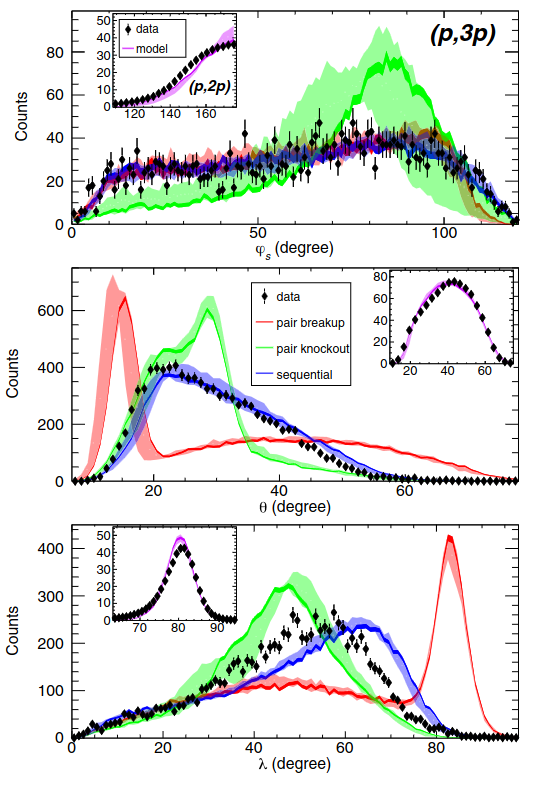}
  \caption{Angular distributions of $^{81}$Ga$(p,3p)$ reaction. Pair breakup, pair knockout, and sequential model calculations are also shown in red, green, and blue bands, respectively. The figure is taken from Ref.~\cite{Frotscher20}.}
  \label{fig:mechanism-p3p}
\end{figure}

The sequential $(p,p2N)$ reaction is also of interest, in addition to the pair knockout process.
The two-proton knockout reaction, $(p,3p)$, was measured from neutron-rich nuclei at $\sim 250 A$~MeV in inverse kinematics~\cite{Frotscher20}.
In this experiment, the angular distribution of the three emitted protons was measured for the first time.
Figure~\ref{fig:mechanism-p3p} shows the angular dependence of the measured $(p,3p)$ events as a function of the projected angle $\varphi_s$, the beam angle $\theta$, and the angle between the scattered protons $\lambda$.
See Fig.~1 of Ref.~\cite{Frotscher20} for the definition of $\varphi_s$, $\theta$ and $\lambda$.
The data were compared with three models: (1) Sequential (2) Pair breakup and (3) Pair knockout.
(1) Sequential process assumes that independent $p$-$p$ binary collision occurs twice sequentially. 
(2) Pair breakup assumes a correlated $p$-$p$ pair inside the nucleus and the center-of-mass of the pair is at rest in the nucleus rest frame. 
The target proton only interacts with one proton of the proton pair and the $p$-$p$ binary collision occurs.
The other proton of the pair, which was a spectator, is emitted with its initial momentum, which the proton had in the proton pair. 
(3) Pair knockout assumes that the center-of-mass of the correlated proton pair is knocked out by the target proton in a similar manner as in the naive $(p,pd)$ reaction.
See Ref.~\cite{Frotscher20} for a detailed explanation.
As shown in Fig.~\ref{fig:mechanism-p3p}, all the angular distributions support the dominance of the sequential process, though the $\lambda$ dependence might suggest a small portion of the pair knockout contribution.

Analyses by more sophisticated reaction theory will give us more insight into the nature of $(p,3p)$ reaction, as well as $(p,ppn)$ and $(p,pnn)$.
It is also interesting if one can select one process out of three possible mechanisms, sequential, pair breakup, and pair knockout by choosing proper kinematics of the $(p,p2N)$ reaction since populated states and their yield by $(p,p2N)$ reaction should depend on the reaction mechanisms.
Depending on the reaction mechanism of $(p,p2N)$ reaction, it is expected to exhibit a different selectivity on the populated state of the residue compared to the $(p,pN)$ reaction, as suggested in Ref.~\cite{Taniuchi19}.
CDCCIA framework discussed in Sec.~\ref{sec:cdccia} will be one of the most suitable theories for describing the pair knockout type $(p,3p)$ process.

\section{Summary}
SEASTAR project and MINOS have revealed important aspects of the nuclear shell structure and its evolution in neutron-rich nuclei.
The simplicity of the reaction mechanism is one of the greatest advantages in understanding and interpreting the knockout reaction observables and their connection to the nuclear structure.
The momentum distribution of the residual nuclei reflects the momentum distribution of the struck nucleon inside the nucleus and thus revealing its single-particle orbitals.
The magnitude of the cross section corresponds to the SF.
Beyond the standard $(p,pN)$ reaction at hundreds of MeV, new physics directions have been discussed in this review.

The nucleon-nucleon correlations play a key role in the two-nucleon pair knockout reactions.
Recently, a combination of the DWIA and CDCC framework, CDCCIA, has been proposed to incorporate the $NN$ correlation properly in the description of the pair-knockout-type $(p,p2N)$ reactions.
At this stage, the CDCCIA framework is applicable only to $(p,pd)$ reaction, but it will shortly be extended to $(p,ppn)$, $(p,3p)$ and $(p,pnn)$ reactions.
The $NN$ correlation inside a nucleus will be directly connected to $(p,p2N)$ observables.

As shown in Fig.~\ref{fig:mechanism-CDCCIA}, the description of the $(p,pd)$ and $(p,ppn)$ reactions includes all relevant reaction paths, such as the breakup and reformation effects on the $p$-$n$ pair from both the elementary process and the final-state interaction.
Although the $p$-$p$ and $n$-$n$ pair do not have the bound state, the continuum-continuum coupling is taken into account similar to the $p$-$n$ pair knockout case.
An comprehensive description covering pair knockout, pair breakup, and sequential $(p,p2N)$ reactions will be necessary for understanding the $(p,p2N)$ reaction mechanism.
It is also important to understand the selectivity for the populated states of the reaction residue by the $(p,p2N)$ reactions, as this selectivity depends on the underlying reaction mechanism.

The $4N$ correlation and the $\alpha$ formation have been of great interest since the beginning of nuclear physics. 
Historically, the $\alpha$ formation has been intensively studied at both ends of the nuclear chart, in light nuclei and as $\alpha$ preformation in the $\alpha$-decaying nuclei.  
Recently, the search for the $\alpha$ formation has been extended to the medium-mass nuclei and the $\alpha$ formation on the surface of Sn isotopes was observed. 
A systematic understanding of the $\alpha$ formation in unstable nuclei from light to heavy mass regions is one of the goals of ongoing ONOKORO project.

This review is based on work presented at the symposium ``Direct Reactions and Spectroscopy with Hydrogen Targets: Past 10 years at the RIBF and Future Prospects'', 31st July to 4th August 2023. 
The authors are grateful for the support by IoP and the JSPS London Symposium and Seminar Scheme (award no: JBUK028) for the symposium.

\newpage
\let\doi\relax
%without this code before the command "\begin{thebibliography}{}" , an error will be %flagged. When the bibliography is provided as a separate .bib file, then this code %should be placed above the commands "\bibliographystyle{}" and "\bibliography{}" %inside the main TeX file. 
\bibliography{ref_normal}
\bibliographystyle{ptephy}

\end{document}